\documentclass[12pt]{article}

\newcommand{\sgn}{{\mathop{\textrm{sgn}}}}

\usepackage{graphicx} 
\newcommand{\eqref}[1]{(\ref{#1})}

\renewcommand{\d}{\partial}
 
\newcommand{\nn}{\nonumber\\}
 
\newcommand{\x}{{\bf x}}
\newcommand{\s}{{\bf s}}
 
\newcommand{\ph}{\varphi}
\newcommand{\rh}{\varrho} 

\newcommand{\exv}[1]{\left\langle{#1}\right\rangle}
\newcommand{\exvs}[1]{\langle{#1}\rangle}
\newcommand{\ep}{\varepsilon} 

\newcommand{\q}{{\bf q}} 
\renewcommand{\k}{{\bf k}}

\newcommand{\Tr}{\mathop{\textrm{Tr}}}
\renewcommand{\Im}{\,\textrm{Im}\,}
\renewcommand{\Re}{\,\textrm{Re}\,}

\renewcommand{\c}[1]{{\cal{#1}}}

\newlength{\szovszel}\newlength{\szovmag}
\newcommand*{\horsl}[2][0.75]{\mbox{%
    \settowidth{\szovszel}{\ensuremath{#2}}%
    \settoheight{\szovmag}{\ensuremath{#2}}%
    \ensuremath{#2}%
    \hspace{-#1\szovszel}\rule[0.8\szovmag]{0.8\szovszel}{0.3pt}}}
\newcommand{\dd}{{\horsl[0.6]{d}}}

\begin{document}
\pagestyle{empty}
\begin{flushright}
  CERN-TH/2002-150\\
  hep-ph/0207085
\end{flushright}
\vspace*{5mm}
\begin{center}
  {\bf GENERALIZED BOLTZMANN EQUATIONS FOR\\ ON-SHELL PARTICLE
    PRODUCTION IN A HOT PLASMA} \vspace*{0.8cm}
  
  A. Jakov\'ac\footnote{e-mail: Antal.Jakovac@cern.ch}\\
  
  \vspace{0.3cm} {\em Theory Division, CERN, CH-1211 Geneva 23,
    Switzerland}
  
  \vspace*{2cm} {\bf ABSTRACT}
\end{center} 
\vspace*{5mm}

\noindent
A novel refinement of the conventional treatment of Kadanoff--Baym
equations is suggested. Besides the Boltzmann equation another
differential equation is used for calculating the evolution of the
non-equilibrium two-point function. Although it was usually
interpreted as a constraint on the solution of the Boltzmann equation,
we argue that its dynamics is relevant to the determination and
resummation of the particle production cut contributions. The
differential equation for this new contribution is illustrated in the
example of the cubic scalar model. The analogue of the relaxation time
approximation is suggested. It results in the shift of the threshold
location and in smearing out of the non-analytic threshold behaviour
of the spectral function.  Possible consequences for the dilepton
production are discussed.

\vfill
\begin{flushleft} CERN-TH/2002-150\\ July 5, 2002
\end{flushleft}
%\vspace*{0.5cm}
\eject

\setcounter{page}{1} \pagestyle{plain}

\section{Introduction}
The real photon production and dilepton production in a hot plasma are
intensively discussed subjects of the current literature. One aspect
of this discussion is the behaviour of the production rate near zero
invariant mass of the lepton pair. Numerical simulations
\cite{Karschetal} find a vanishing rate, while higher loop
perturbative \cite{AltherrAurenche} and HTL-improved
\cite{BraatenPisarski} calculations find diverging result in this
limit. The possibility that the two-loop order enhancement may be a
consequence of infrared (IR) singularities was studied in recent
papers \cite{Aurenche}, where also an effective description was
suggested to overcome the problem. In \cite{AMY} a resummation method
was worked out for ladder diagrams, and it was applied to the photon
and gluon production rate.

Resummation in perturbation theory becomes unavoidable, in general,
when we encounter IR divergences that enhance the contribution of
higher order loops, thus overriding the coupling constant suppression.
A well known example is the on-mass-shell IR divergence of the
propagator due to a radiative mass shift. This divergence is cured by
resumming 1PI diagrams with the help of Schwinger--Dyson (SD)
equations.  This example also illustrates that resummation may be
effective only for some specific momentum range, while in other
regions its effect might remain negligible. From this point of view,
also finite-temperature Debye screening is a consequence of IR
divergences.

In non-equilibrium systems there appear new sources of IR divergences.
When working in real time one recognizes that higher loops may give
contributions that grow as some power of the elapsed time ($\sim
t^{\alpha n}$ where $n$ is the loop order). This would clearly
restrict the applicability of perturbation theory to rather short
times, so we need a resummation. It turns out \cite{BVW} that
resummation leads to Boltzmann equations describing the temporal
evolution of the propagator. In Fourier space the corresponding
phenomenon is the appearance of pinch singularities
\cite{LandsvanWeert,pinch-sing} in two-particle intermediate states
with vanishing momentum difference. The resummation of these
divergences, which involves ladder diagrams, leads, again, to
Boltzmann equations \cite{AJ}.

In this paper we shall argue that there are further momentum ranges
where we can observe pinch singularities. When these momentum ranges
are relevant, we have to perform appropriate resummation also there.
To see the origin of these divergences we have to analyse the
corresponding pinch singularities, as will be done in
Section~\ref{sec:highpinch}. The core effect of pinch singularities is
that in a pair of propagators one of them can force the other to its
mass shell; for instance in $G_R(p) G_A(q) = G_R(p) i\rh(q) + G_R(p)
G_R(q)$ the first term yields the singular effect, the second turns
out to be regular. If we parametrise the momenta carried by the
particles as $p=Q+ K/2$ and $q=Q-K/2$, then we can find for the
contributions two formally equivalent expressions:
\begin{equation}
  G_R(p) \rh(q) \to \frac{\delta((Q-K/2)^2-m^2)}{(Q+K/2)^2-m^2} =
  \frac{\delta((Q-K/2)^2-m^2)}{2QK} =\frac12
  \frac{\delta((Q-K/2)^2-m^2)}{Q^2+K^2/4-m^2}.
\end{equation}
(The Landau prescription is omitted to simplify this introductory
argument.) Thus the product of the propagators becomes divergent
either when $2QK=0$ (this case of the resummation leads to the
Boltzmann equation), \emph{or} when $Q^2+K^2/4-m^2=0$. This second is
the new kinematical situation studied in the present paper.

After identifying the IR divergences, we have to design an appropriate
resummation scheme for them. In order to perform ladder resummation we
can use the linearized Kadanof--Baym (KB) equations \cite{AJ}. These
come from a procedure (for details see Sections~\ref{sec:SD}
and~\ref{sec:highpinch}), which consists of writing two SD equations
for each of the Wigner--Fourier-transformed propagators, next taking
their sum and difference, and finally linearizing them. They read,
schematically:
\begin{equation}
    2QK\bar G(Q,K) = (\c A_- * \bar G)(Q,K),\qquad
    2(Q^2+\frac{K^2}4-m^2)\bar G(Q,K) = (\c A_+ * \bar G)(Q,K),
\end{equation}
where $\c A_\pm$ denote the linear operators acting on the
propagators. The equations are diagonal in $K$, which is the momentum
associated with the average coordinate of the propagator. The
coefficient of $\bar G$ vanishes on the left-hand side in regions of
pinch singularity, while the expression is finite on the right-hand
side. The perturbative solution diverges at every order in any such
regime, but the complete solution of the differential equations is
regular. This is the resummation realized with the help of SD
equations.  Using other resummation techniques will give the same
result, at least to leading order of the pinch-singular terms
\cite{AJ}.

Our new strategy is to suggest that in a general kinematical region
where pinch singularities occur we have to solve simultaneously
\emph{both} KB equations for the propagator. It differs from the usual
approach, where one treats the sum of the SD equations, for small $K$,
as a constraint, which effectively puts $Q$ on the mass shell. But how
can one use two different solutions for the same propagator?  As will
be shown in Section~\ref{sec:oneloop}, at one loop level these
contributions are relevant to different kinematical regions: in
Boltzmann equations we must have $k_0<2q_0$, in the other case (called
normal cut regime in the following) $k_0>2q_0$ must be true.  At one
loop the exactly known complete solution is thus the sum of the
respective KB equations. We propose to maintain this additivity
feature also for the resummed propagator. We expect that the
kinematical complementarity of the two solutions is a good
approximation while the two-particle threshold is well separated from
the regime where Landau-damping dominates.

While in our approach no constraint equations play any role, in the
leading order pinch-singular terms we can solve the SD equations with
an ansatz implying the missing constraints (see
Sections~\ref{sec:boltzeq} and ~\ref{sec:normalcut}). In this way the
usual solutions for the Boltzmann equations survive, we just have
supplementary contributions to the propagator from the complete
solution.

Computing expectation values of composite operators involves
integration over $Q$, which is the momentum associated with the
relative coordinate. After this integration, more insight is offered
into the physics of the different contributions. At one loop order the
discontinuity of the Boltzmann term $\sim1/(QK)$ gives a non-zero
result for $K^2<0$ -- this is the Landau damping regime.  The
discontinuity of the other contribution $\sim1/(Q^2+K^2/4-m^2)$ is
non-zero when $K^2> 4m^2$ where $2m$ is the threshold value for
particle production -- this is a normal cut regime. The solution of
the Boltzmann equation in the relaxation time approximation generates
a finite imaginary part to $k_0$; this is why a quantity, conserved at
tree level, can decay exponentially due to radiative effects. In a
kind of similar relaxation time approximation, higher loop pinch
singularities provide imaginary \emph{and} real contributions to the
denominator $Q^2+K^2/4-m^2$, which results in a shift of the threshold
location and in a ``smoothing'' of the threshold singularity.

In this paper, the effects described above in qualitative terms will
be demonstrated in the cubic scalar model. Beyond its simplicity it
shows some similarity with fermion--gauge systems as far as the
perturbation theory is concerned. We choose a simple observable,
$\exv{[\ph^2(x),\ph^2(0)]}$; this is the simplified analogue of
$\exv{[j_\mu(x),j^\mu(0)]}$, which plays an important role in the
theory of dilepton production \cite{dilepton}. In
Section~\ref{sec:model} we introduce the model and our observable. In
Section~\ref{sec:oneloop} we perform a thorough one-loop calculation,
identify the different kinematical regions, and examine the way SD
equations and KB equations reproduce the one-loop result. In
Section~\ref{sec:highpinch} we turn to higher loops. We show the
origin of pinch singularities, and apply KB equations to resum these
singularities in different kinematic regions. We also solve the
corresponding equations in relaxation time approximation. In
Section~\ref{sec:fermcomments} we make some comments on the relevance
of our results to the dilepton production in fermion--gauge systems.
Section~\ref{sec:conclusion} then closes the paper by presenting our
conclusions and an outlook.

\section{The model}
\label{sec:model}

The cubic scalar model has the Lagrangian
\begin{equation}
  \c L = \frac12 (\d\ph)^2 - \frac12 m^2\ph^2 -\frac h6\ph^3.
\label{lagra}
\end{equation}
We are interested in the equilibrium expectation value of the
composite spectral function:
\begin{equation}
  \exvs{[\ph^2(x),\ph^2(0)]}\biggr|_{1PI,\exv\ph=0}.
\end{equation}
This quantity corresponds to the expectation value
$\exvs{[j_\mu(x),j^\mu(0)]}$ in the fermion--gauge systems, which
plays a central role in dilepton production \cite{dilepton}. The
subscripts 1PI and $\exv\ph=0$ will be omitted in the following. The
quantity of phenomenological interest is the discontinuity of $iR(K)$,
which is the Fourier transform of
\begin{equation}
  iR(x)=\Theta(x_0)\exvs{[\ph^2(x),\ph^2(0)]}.
\end{equation}
$R(x)$, on the other hand, is the linear response function of
$\exv{\ph^2(x)}$ to the mass modification of the Lagrangian
characterized by the linear operator $\c P$:
\begin{equation}
  \c L_{\c P} = \c L - \frac12 \c P \ph^2 \qquad\Rightarrow\qquad
  \delta \exv{\ph^2}(K) =\frac12 R(K) \c P(K) + \dots,
\end{equation}
where $\delta f = f - f|_{\c P=0}$ for any quantity. We will define
$\exvs{\ph^2(x)}$ via point splitting; in the Keldysh and in the R/A
formalisms (cf. Appendix~\ref{sec:formalism}) we should compute
\begin{equation}
  \exv{\ph^2(x)} = \frac12\lim_{y\to x}\left[\exv{\ph(x)\ph(y)} +
  \exv{\ph(y)\ph(x)} \right] = \frac12\lim_{y\to x} \left[iG_{21}(x,y)
  + iG_{12}(x,y)\right] = \lim_{y\to x}i G_{rr}(x,y).
\end{equation}
In Fourier space
\begin{equation}
  \exv{\ph^2}(K) = \int\!dx\, e^{iKx} \lim_{y\to x} \int\!\dd p\dd q\,
  e^{-ipx + iqy} i G_{rr}(p,q)  = \int\!\dd Q\,i\bar G_{rr}(Q,K),
  \label{ph2andGrr}
\end{equation}
($\bar G(Q,K)$ denotes the Wigner--Fourier transform of $G(x,y)$, which
has the usual two-variable Fourier-transform $G(p,q)$, cf.
Appendix~\ref{sec:formalism})). We introduce the quantity $R(Q,K)$
defined through the relation
\begin{equation}
  \delta i \bar G_{rr}(Q,K) = \frac12 R(Q,K) \c P(K)
  \qquad\Rightarrow\qquad  R(K) = \int\!\dd Q\, R(Q,K).
\end{equation}

\section{One-loop perturbation theory}
\label{sec:oneloop}

In leading order the only interaction we have to take into account is
$-\c P\ph^2/2$. In the Keldysh and R/A formalism (cf.
Appendix~\ref{sec:formalism}) this piece of the Lagrangian has the
form
\begin{equation}
  -\frac12 \c P \ph^2 \longrightarrow -\frac 12 \c P \left(\ph_1^2
   -\ph_2^2\right) = -\c P \ph_r\ph_a.
 \label{intLag}
\end{equation}
Linear response to $\c P$ yields
\begin{equation}
  \delta \bar G_{rr}(Q,K) = \left[ {\bar G}^0_{ra}(Q+\frac K2) {\bar
  G}^0_{rr}(Q -\frac K2) + {\bar G}^0_{ar}(Q-\frac K2) {\bar
  G}^0_{rr}(Q +\frac K2) \right] \c P(K),
\end{equation}
where $G_{ab}^0$ means free equilibrium propagators. The coefficient
of $\c P$ identifies $-iR(Q,K)/2$. Its discontinuity reads
\begin{equation}
  \mathop{\mathrm{Disc}}_{k_0} iR(Q,K) = 2\left[\rh^0(Q+\frac K2)
  i{\bar G}^0_{rr}(Q -\frac K2) - \rh^0(Q-\frac K2) i{\bar G}^0_{rr}(Q
  +\frac K2)\right].
\end{equation}
Inserting the free equilibrium expressions ($\omega^2_\q = \q^2+m^2$)
\begin{equation}
  \rh^0(Q) = 2\pi \sgn(q_0) \delta(q_0^2-\omega_\q^2),\qquad
  iG_{rr}^0(Q) = 2\pi \sgn(q_0) \left(\frac12 + n(q_0)\right)
  \delta(q_0^2-\omega_\q^2)
\end{equation}
into this equation yields
\begin{equation}
  \mathop{\mathrm{Disc}}_{k_0} iR(Q,K) = 8\pi^2
  \sgn(q_0+\frac{k_0}2) \sgn(q_0 - \frac{k_0}2)(n_- - n_+)
  \delta((Q+\frac K2)^2-m^2) \delta((Q-\frac K2)^2-m^2),
  \label{DiscR}
\end{equation}
where $n_\pm = n(q_0\pm k_0/2)$.

We first analyse the product of the delta functions (for details,see
Appendix~\ref{sec:compdet}). One can write them in various forms:
\begin{eqnarray}
  D:=&& \delta((Q+\frac K2)^2-m^2) \delta((Q-\frac K2)^2-m^2) =
  \delta(2QK) \delta(Q^2+\frac{K^2}4 - m^2) \nn 
  =&& \frac1{2\,{\c Z}\,|k_0^2-k^2x^2|} \delta(q_0-\frac{kx}2
  \sgn(k_0){\c Z}) \delta(q-\frac{|k_0|}2{\c Z}),
  \label{proddelta}
\end{eqnarray}
where $\c Z=\sqrt{(K^2-4m^2)/({k_0^2-k^2x^2})}$, which is real when
$K^2>4m^2$ or $K^2<0$ (see Appendix~\ref{sec:compdet}).

The last form carries an important message concerning the relative
magnitude of $q_0$ and $k_0$. For $K^2<0$ it requires that
$|q_0|>|k_0|/2$, while for $K^2>4m^2$ we find $|q_0|< k/2 <|k_0|/2$
(see Appendix~\ref{sec:compdet}). It has the consequence that in the
present circumstances one can write:
\begin{equation}
  \sgn(q_0\pm \frac{k_0}2) = \Theta(K^2-4m^2) (\pm) \sgn(k_0)
  + \Theta(-K^2) \sgn(q_0),
\end{equation}
which implies, when inserting back into \eqref{DiscR}, that
\begin{eqnarray}
  && \mathop{\mathrm{Disc}}_{k_0} iR(Q,K) =\frac{4 \pi^2}
  {\sqrt{(k_0^2-k^2x^2) (K^2-4m^2)}}\delta(q_0-\frac{kx}2
  \sgn(k_0){\c Z}) \delta(q-\frac{|k_0|}2{\c Z}) \times\nn&&
  \left[\Theta(K^2-4m^2) \left(1+ n(\frac{k_0}2-q_0)+
      n(\frac{k_0}2+q_0)\right) + \Theta(-K^2) \left(n(q_0-\frac{k_0}2) -
      n(q_0+\frac{k_0}2)\right) \right].
  \label{DiscR:qkform}
\end{eqnarray}

The $Q$ integration is performed using this form (see
Appendix~\ref{sec:compdet}), and one finds
\begin{eqnarray}
  \int\!\dd Q\, \mathop{\mathrm{Disc}}_{k_0} iR(Q,K) =
  \frac{\sgn(k_0)}{4\pi k} \biggl[ && \Theta(K^2-4m^2)
  \int\limits_{-k{\c Z}_1/2}^{k{\c Z}_1/2}\!dy\, \left( 1 +
    n(\frac{|k_0|}2-y) +n(\frac{|k_0|}2+y)\right) +\nn&&  2\Theta(-K^2)
  \int\limits_{k{\c Z}_1/2}^\infty\! dy\, \left( n(y-\frac{|k_0|}2) -
    n(y+\frac{|k_0|}2) \right)\biggr],
  \label{DiscR:Qintform}
\end{eqnarray}
which agrees with the results of earlier calculations
\cite{oneloopcalc}.

The limiting case usually investigated is that the spatial variation
of the external source is small, i.e. $k\ll m, T$. Then, in the first
term of eq.~\eqref{DiscR:Qintform}, the variation of $y$ is of order
$k$, thus $y\ll|k_0|$ and we can neglect the $y$ dependence of the
integrand. In the second term $k\ge|k_0|$, so the small-$k$
approximation yields ${\c Z}_1\approx 2m/|K|\gg1$. Then $y\gg k\ge
k_0$ and we can perform an expansion in power series with respect to
$k_0$. Finally we obtain
\begin{equation}
  \int\!\dd Q\, \mathop{\mathrm{Disc}}_{k_0} iR(Q,K) \approx
  \frac1{4\pi} \biggl[ \Theta(K^2-4m^2) \sqrt{1-\frac{4m^2}{K^2}}
  \left( 1 + 2 n(\frac{k_0}2)\right) +  2\Theta(-K^2) \frac{k_0}k
  n(\frac{km}{|K|})\biggr].
  \label{appDiscR}
\end{equation}

\subsection{The Landau damping region}

The approximate expression \eqref{appDiscR} can be derived without the
detailed computation outlined above. If $K^2<0$ and $k\ll m,\,T$, the
plasma excitations behave as on-mass-shall particles because $Q^2 =
m^2-K^2/4\approx m^2$. At high temperatures all components of $Q$ are
of the order of the temperature. In this regime it is a good
approximation to neglect $K$ in \eqref{DiscR} and \eqref{proddelta}
whenever it is possible, and to perform an expansion in power series
with respect to the components of $K$ when it is necessary. This
yields
\begin{equation}
  \mathop{\mathrm{Disc}}_{k_0} iR(Q,K) \approx -8\pi^2k_0
  \,\frac{dn(q_0)} {dq_0} \delta(2QK) \delta(Q^2-m^2) = -8\pi^2 k_0
  \frac{dn}{d\omega} \delta(2QK) \delta(Q^2-m^2),
\end{equation}
since $dn/dq_0$ is even in $q_0$. After the $Q$ integration one
readily reproduces the Landau damping piece of \eqref{appDiscR}.

Although this is a straightforward consequence of our approximations,
this result is still a bit surprising. The assumption we made is
quite plausible: the variation of the background field $K$ is small as
compared to the typical scale of the loop momentum, which is $T$ at
high temperatures. This was implemented by requiring $Q\gg K$ for all
components. This assumption, however, just reproduces the Landau cut
region, while the normal cut is missing.

\subsection{The normal cut region}

Let us examine where we have lost the normal cut, and what are the
appropriate assumptions that lead to its approximate form in
\eqref{appDiscR}, in a similarly direct way as the Landau cut is
reproduced.

The answer is easy to find when we recall the earlier statement of the
exact analysis that in the $K^2>4m^2$ regime $|q_0|<k/2<|k_0|/2$: that
means that the assumption $Q\gg K$ is not valid for all components. In
the normal cut regime the time-like component of the loop momentum is
small, even smaller than the spatial variation of the background
field.

This observation suggests a complementary approximation scheme: we
neglect the (spatial) $k$ dependence, not assuming, however, that
$k_0<k$.  In the product of the delta functions \eqref{proddelta} we
can then write
\begin{equation}
  D = \delta(2QK)\delta(Q^2+\frac{K^2}4 -m^2) \longrightarrow
  \delta(2q_0k_0) \delta(Q^2+\frac{k_0^2}4 -m^2) = \frac1{2|k_0|}
  \delta(q_0) \delta(\frac{k_0^2}4 -m^2-q^2).
\end{equation}
Therefore, in the leading order, we can assume $q_0\approx0$ as well.
We also see that we immediately obtain $k_0^2>4m^2$ from the second
delta function. The approximation for the discontinuity of $R$ then
reads
\begin{equation}
  \mathop{\mathrm{Disc}}_{k_0} iR(Q,K) \approx \frac{4\pi^2}{|k_0|}
  \left(1+2 n(\frac{k_0}2) \right) \delta(q_0) \delta(\frac{k_0^2}4
  -m^2-q^2).
\end{equation}
This expression reproduces, after $Q$ integration, the normal cut part
of the approximated result \eqref{appDiscR}.

It is interesting to observe that the approximations leading to Landau
and normal cuts concern different factors of $D$ in \eqref{proddelta}:
using $\delta(Q^2+K^2/4-m^2)$ for small $K$ yields that $Q$ is on the
mass shell, while using $\delta(2QK)$ for small $k$ yields
$q_0\approx0$. While both assumptions seem to be well justified,
neither of them reproduces the complete discontinuity of $R$, just
complementary parts of them. To understand this phenomenon better, we
now turn to the Schwinger--Dyson equations.

\subsection{Schwinger--Dyson equations}
\label{sec:SD}

The general form of SD equations in Keldysh and R/A formalism can be
found in Appendix~\ref{sec:formalism}. At one loop perturbation theory
we need the self-energies coming from the interaction Lagrangian
\eqref{intLag}:
\begin{equation}
  \Pi_{ar}(Q,K)=\Pi_{ra}(Q,K)=\c P(K),\qquad \Pi_{aa}=0,
\end{equation}
which has to be inserted into \eqref{WigSD} of
Appendix~\ref{sec:formalism}. For a linear response we insert free
propagators into its right-hand side. We are interested in the $rr$
component (cf. \eqref{ph2andGrr})
\begin{equation}
  \left((Q\pm\frac K2)^2-m^2\right) \bar G_{rr}(Q,K) = \c P(K)
  {\bar G}^0_{rr}(Q\mp\frac K2).
\end{equation}
The two equations, like the general ones, are related by complex
conjugation and $K\to-K$ transformation.

Taking the sum and the difference of the two equations we find
\begin{eqnarray}
  \label{sumdiff1}
  &&\left(Q^2 +\frac{K^2}4-m^2\right) \bar G_{rr}(Q,K) = \c P(K)
  \frac{{\bar G}^0_{rr}(Q+\frac K2) +{\bar G}^0_{rr}(Q-\frac K2)}2 \\
  \label{sumdiff2}
  &&2QK \bar G_{rr}(Q,K) = \c P(K)\left[ {\bar G}^0_{rr}(Q-\frac K2)
    -{\bar G}^0_{rr}(Q+\frac K2) \right].
\end{eqnarray}
This yields two solutions for $\bar G_{rr}(Q,K)$. Since the two SD
equations are not independent, the two solutions should imply the same
physics. To see this we recall that ${\bar G}^0_{rr}(Q)\sim \delta
(Q^2-m^2)$, therefore
\begin{equation}
  \frac{{\bar G}^0_{rr}(Q+\frac K2)}{Q^2 +\frac{K^2}4-m^2} =
  -\frac{{\bar G}^0_{rr}(Q+\frac K2)}{2QK}.
\end{equation}
On the other hand the equivalence is not complete, since we lose
something from the analytical structure. If we treat these equations
as differential equations with initial condition (i.e. we apply the
Landau prescription), and we take the imaginary part, the two
equations give different results. We use the following form of the
free $rr$ propagator ($\omega^2=\q^2+m^2$):
\begin{equation}
  i{\bar G}^0_{rr}(Q) = 2\pi \left(\frac 12 + n(\omega)\right)
  \delta(Q^2-m^2) 
\end{equation}
and write for the solution of \eqref{sumdiff1} and \eqref{sumdiff2}
\begin{eqnarray}
  && \mathop{\mathrm{Disc}}_{k_0} i\bar G_{rr}(Q,K) = 4\pi^2
  \c P(K) \sgn(k_0)\left(1 + n(\omega_+) + n(\omega_-)\right)
  \delta(Q^2 +\frac{K^2}4-m^2)  \delta(2QK),\nn
  && \mathop{\mathrm{Disc}}_{k_0} i\bar G_{rr}(Q,K) = 4\pi^2
  \c P(K) \sgn(q_0)\left(n(\omega_-) - n(\omega_+)\right)
  \delta(Q^2 +\frac{K^2}4-m^2)  \delta(2QK),
\end{eqnarray}
respectively, where $\omega_\pm = \omega(\q\pm\k/2)$. Here the first
form corresponds to the normal cut regime, the second to the
Landau-cut regime. Although formally there seems to be nothing that
could prohibit applying, for example, the second equation to the
$K^2>4m^2$ region, the result will be incorrect.

If we compare these formulae to the ones coming from direct
perturbation theory (see \eqref{DiscR} and \eqref{proddelta}) we can
see that the difference is that that in the exact formula there stands
$\sgn(q_0\pm k_0/2)$, which is substituted by $\sgn(k_0)$ and
$\sgn(q_0)$ in the first and the second form of the above equations,
respectively. The reason for this is simple: when we take for example
the difference of the two equations, $K^2$ disappears, and so the
corresponding $ik_0\ep$ will be missing in the later calculations,
too. Therefore we implicitly assume that the $2iq_0\ep$ term (coming
from $2QK$) dominates it: i.e.  $q_0>k_0/2$. This implicit assumption
prevents the final result to be applicable everywhere.

Usually one considers this pair of equations as a differential
equation (\eqref{sumdiff2}, the Boltzmann equation) and, assuming $K
\ll Q$, a constraint \eqref{sumdiff1}. But this strategy, as we have
seen, misses the normal cut part from the very beginning. From our
analysis a slightly different picture seems to emerge: here both
equations are differential equations to be solved, and the complete
solution is the sum of the two solutions.  We propose to apply this
strategy also for the non-perturbative, non-linear response regime: we
should solve not just the equation coming from the difference of the
pair of SD equations, i.e. the Boltzmann equation. We have to treat
the sum of the two equations as a differential equation, since this is
the approach that makes explicit where the potentially important
(especially in the zero-mass case) normal cut contribution comes from.
The complete solution then can be composed as the sum of these partial
solutions.

But where are the constraint equations? In our case the right-hand
side, and so the solution was explicit, there was no need to add
another equation. At higher loops or in the non-linear response
regime, as we will see in the next section, concentrating on the IR
sensitive pinch-singular region, we can still have a consistent ansatz
that satisfies the (by now missing) constraint equation.

As a last remark in this section we emphasize the importance of the
cut contributions in free systems. Let us consider the evolution of
$\int d^3\x\exvs{\ph^2}(\x,k_0) = \exvs{\ph^2}(\k=0,k_0)$ in a free
system, without any background field, but with specified initial
conditions. The relevant equations are the following (cf.
\eqref{sumdiff1}, \eqref{sumdiff2}):
\begin{equation}
  2q_0k_0 \bar G_{rr}(Q,k_0) =0,\qquad \left(Q^2
  +\frac{k_0^2}4-m^2\right) \bar G_{rr}(Q,K) = 0.
\end{equation}
Taking into account only the first equation we would conclude that
$\bar G_{rr}(Q,t)$ is constant, and thus $\exvs{\ph^2}(\k=0,t)$ is
constant, too. This would mean that $\exvs{\ph^2}(\k=0,t)$ is a
conserved quantity -- which it is not, in reality. Analysing the
second equation (cf. \cite{oneloopcalc,background-time}), we find that
it yields a power law time dependence, that is $\exvs{\ph^2}(\k=0,t)$
goes to zero, with some power of time. This again shows that the
Boltzmann equation, alone, cannot provide the complete treatment even
in the simplest case.

\section{Higher loops and pinch singularities}
\label{sec:highpinch}

Higher loops are, of course, suppressed by powers of the coupling
constant; therefore they become important only if the one-loop result
is zero or there are IR divergences. The imaginary part of the
one-loop result, as we have seen, is zero if $0<K^2<4m^2$; at two
loops, however, this restriction is relaxed and non-zero imaginary part
is obtained for all $K$. As far as the IR divergences are concerned,
in our case the pinch singularities \cite{LandsvanWeert,pinch-sing}
are the primary sources of them.

If the product of a retarded and an advanced propagator appears in
some objects with the same momentum, then the poles of the propagators
pinch the real axis as $\ep\to0$ and yield a non-regularizable double
pole. If the two momenta are not equal, we expect a singular behaviour
of the object as the difference of the momenta vanishes. Using the
fact that $G_RG_R$ is not singular and that $i(G_R-G_A)=\rh$, we can
write
\begin{equation}
  G_R(Q+\frac K2)G_A(Q - \frac K2) = G_R(Q+\frac K2)\,i\rh(Q-\frac K2)
  - \mbox{\textrm{non-singular}}.
\end{equation}
If $K\to 0$ the free spectral function $\rh(Q)\sim \delta(Q^2-m^2)$
puts $G_R$ on-shell: this is the core effect of the pinch
singularities. On the other hand, $G_{rr}\sim \rh$, thus similar
singularities appear when we multiply $G_{R,A}$ and $G_{rr}$. We have
to treat this case in complete analogy with the $G_RG_A$ product.

Pinch singularities appear in all higher order diagrams containing
two-particle intermediate states \cite{AJ}; to see the net effect we
have to sum up all the diagrams of this type. In the linear response
case, these diagrams form ladders.

Which are the kinematic ranges where we have to count with pinch
singularities? From the above generic form
\begin{equation}
  G_R(Q+\frac K2)\,i\rh(Q-\frac K2) = \frac{ i\rh(Q-\frac K2) } {2QK}
  = \frac12\, \frac{ i\rh(Q-\frac K2) } {Q^2+K^2/4-m^2}.
\end{equation}
There are therefore two sensitive regimes: the vanishing $2QK$ or
the vanishing $Q^2+K^2/4-m^2$ regime.

To resum the pinch singular diagrams we can use any valid equation
where the left-hand side is proportional to $2QK$ or $Q^2+K^2/4-m^2$,
respectively, and the right-hand side is non-vanishing in the $2QK=0$
or $Q^2+K^2/4-m^2=0$ limit \cite{AJ}; the difference in the exact
choice of the equation just gives a subleading effect. In linear
response theory we need a linearized right-hand side. Exactly of this
form are the linearized SD equations \eqref{WigSD} of
Appendix~\ref{sec:formalism}. Starting from it, and linearizing around
equilibrium in the form
\begin{equation}
  \bar G(Q,K) = G^{eq}(Q) \horsl\delta(K) +g(Q,K),\quad \bar
  \Pi(Q,K)=\Pi^{eq}(Q) \horsl\delta(K)+\pi(Q,K),
\end{equation}
%using equilibrium SD equation, and writing the result in the form of
%sums and differences we find
one finds the following equations for the deviations:
\begin{eqnarray}
  \left(Q^2 + \frac{K^2}4-m^2\right) 2g_{rr} =&& \pi_{aa}
  \left(G^-_{ar}+G^+_{ra}\right) + \Pi^+_{aa}g_{ar} + \Pi^-_{aa}g_{ra}
  +  \nn&& \pi_{ar}G^-_{rr} +\pi_{ra}G^+_{rr}  +
  \left(\Pi^+_{ar}+\Pi^-_{ra}\right) g_{rr}\nn 
  2QK g_{rr} = && \pi_{aa} \left(G^-_{ar} - G^+_{ra}\right) +
  \Pi^+_{aa}g_{ar} - \Pi^-_{aa}g_{ra} +  \nn&& \pi_{ar}G^-_{rr} -
  \pi_{ra}G^+_{rr} + \left(\Pi^+_{ar}-\Pi^-_{ra}\right) g_{rr}\nn  
  \left(Q^2 + \frac{K^2}4-m^2\right) 2g_{ra} = && \pi_{ar}\left(
  G^-_{ra}+G^+_{ra}\right) + \left(\Pi^+_{ar}+\Pi^-_{ar}\right)
  g_{ra}\nn 
  2QK g_{ra} = && \pi_{ar}\left( G^-_{ra}-G^+_{ra}\right) +
  \left(\Pi^+_{ar}-\Pi^-_{ar}\right) g_{ra},
\label{sumanddiff}
\end{eqnarray}
where we have omitted the $(Q,K)$ arguments of the functions and we
have introduced the notations $F^\pm \equiv F^{eq}(Q\pm K/2)$.  The
$\c P$ dependence is hidden in $\pi_{ra}$ and $\pi_{ar}$
\begin{equation}
  \pi_{ra}\to \c P + \pi_{ra},\qquad \pi_{ar} \to \c P + \pi_{ar}.
\end{equation}

The self-energies (apart from the term containing $\c P$) come from
loop diagrams, i.e. they consist of propagator products supplemented
with some integration. In a linear response, all propagators but one
are in equilibrium, so they are functions of space-time coordinate
difference.  Therefore the general structure reads in Fourier space
\begin{equation}
  \pi(p,q) \sim \int \dd k\dd\ell\, \c K(p,-q,k,-\ell) g(k,\ell),
\end{equation}
where we divided the momenta into incoming ($p,k$) and outgoing
($q,\ell$) ones, but this is just a convention in the signs. Since $\c
K $ contains equilibrium propagators only, the momentum is conserved,
and we can write
\begin{equation}
  \pi(p,q) \sim \int\dd\ell\, \c K'(p,-q,\ell) g(p-q+\ell,\ell).
\end{equation}
The Wigner transformed function reads
\begin{equation}
  \bar \pi(Q,K) \sim \int\dd\ell\, \bar{\c K'} (Q,K,\ell) \bar
  g(K+\ell, K)=\int\dd\ell\,\bar{\c K''} (Q,K,\ell)\bar g(\ell, K).
\end{equation}
In general, we have to sum over the R/A indexes, too. From this form
we see that the linearized SD equations are diagonal in $K$.

\subsection{The Boltzmann equations}
\label{sec:boltzeq}

Instead of using the full set of SD equations in \eqref{sumanddiff},
we can make some approximations in order to construct equations that
can be treated more easily.  According to our earlier analysis, we will
concentrate on the summation of pinch singularities, which covers two
regions: the small $QK$ region and the small $Q^2+K^2/4-m^2$ region.
The approximation procedure will be to power-expand everything with
respect to the powers of these small parameters.

We start with the Boltzmann region: here $q_0>k_0/2$ and the
differential equations are the second and fourth of
\eqref{sumanddiff}.  First we examine the equations for $g_{ra}$:
\begin{equation}
  2QK g_{ra} = \pi_{ar}\left( G^-_{ra}-G^+_{ra}\right) +
  \left(\Pi^+_{ar}-\Pi^-_{ar}\right) g_{ra}.
\end{equation}
If $q_0>k_0/2$ and in the small $QK$ case the following approximation
is valid
\begin{equation}
  G^\pm_{ra}= \frac1{Q^2+K^2/4-m^2 \pm QK +iq_0\ep} =
  \frac1{Q^2+K^2/4-m^2+iq_0\ep} +\c O(QK),
\label{gGraapp}
\end{equation}
therefore the difference of these functions is zero up to $\c O(QK)$.
Then the equation for $g_{ra}$ is homogeneous (no mixing with
$g_{rr}$), therefore a valid ansatz is to assume that $g_{ra}\equiv
0$. In a similar way we can assume that $g_{ar}\equiv 0$. This is a
generally accepted approximation: the usual Boltzmann equation never
leads to non-zero $g_{ra}$ and $g_{ar}$.

Having said this, the equation for $g_{rr}$ reads
\begin{equation}
  2QK g_{rr} = \pi_{aa} \left(G^-_{ar} - G^+_{ra}\right) +
  \pi_{ar}G^-_{rr} - \pi_{ra}G^+_{rr} +
  \left(\Pi^+_{ar}-\Pi^-_{ra}\right) g_{rr}.
\end{equation}
We approximate the free $G_{rr}$ for small $2QK$ and $q_0>k_0/2$ as
(here $n_\pm=n(q_0\pm k_0/2)$):
\begin{equation}
  iG^\pm_{rr}=2\pi\sgn(q_0\pm\frac{k_0}2) \left(\frac12+n_\pm \right)
  \delta(Q^2+\frac{K^2}4-m^2 \pm QK) \approx 2\pi \sgn(q_0)
  \left(\frac12+n_\pm \right) \delta(Q^2+\frac{K^2}4-m^2),  
\end{equation}
while the difference between $G_{ra}$ and $G_{ar}$, as we have seen in
\eqref{gGraapp}, reads as
\begin{equation}
  G^-_{ar} - G^+_{ra} = 2\pi i \sgn(q_0) \delta(Q^2+\frac{K^2}4-m^2) +
  \c O(QK).
\end{equation}
Therefore all explicit terms in the equation are proportional to
$\sgn(q_0)\delta(Q^2+K^2/4-m^2)$; a consistent solution, therefore,
comes by writing the ansatz
\begin{equation}
  ig_{rr}(Q,K) = 2\pi\sgn(q_0) \delta(Q^2+\frac{K^2}4-m^2) \delta
  n(Q,K).
\end{equation}
Therefore, without making use of the other equation arising from the
sum of the two original SD equations, we are led to a form consistent
with this constraint. This is now, however, somewhat weaker statement,
it is just an ansatz, and there may exist other solutions that do not
respect this ansatz.

If we are interested in the small $\k$ limit of the result, we can
make other simplifications as well. Since in this region $K^2<0$, it
also means a small $k_0$, and we can thus neglect $K^2$ on the
right-hand side. We find finally
\begin{equation}
    2QK\delta n = -\pi_{aa} + (\pi_{ar}-\pi_{ra}) \left( \frac12 +
  n(q_0)\right) + (\Pi^0_{ar}-\Pi^0_{ra})\delta n.
\end{equation}
This is the linearized Boltzmann equation when we expand $\pi$'s in
terms of $\delta n$.

In our concrete model \eqref{lagra}, using the ansatz, we find
\begin{eqnarray}
  && \pi_{aa} = -ih^2\int\!\dd L\,\rh(L)\rh(Q-L)\left[
    \left(\frac12+n(q_0-\ell_0)\right) \delta n(\ell_0) +\left(\frac12 +
      n(\ell_0)\right) \delta n(q_0-\ell_0) \right]\nn
  && \pi_{ar}-\pi_{ra} = -ih^2 \int\!\dd L\,\rh(L)\rh(Q-L) [\delta
  n(\ell_0) + \delta n(q_0-\ell_0)].
\end{eqnarray}
The equilibrium value of the self energy reads
\begin{equation}
  \Pi^0_{ar}-\Pi^0_{ra} = -ih^2 \int\!\dd L\,\rh(L)\rh(Q-L)
  (1+n(\ell_0) + n(q_0-\ell_0)),
\end{equation}
and thus for the integrand on the right hand side we find
\begin{equation}
  \delta n(\ell_0) (n(q_0)-n(q_0-\ell_0)) + \delta n(q_0-\ell_0)
  (n(q_0)-n(\ell_0)) + \delta n(q_0)(1+n(\ell_0)+n(q_0-\ell_0)),
\end{equation}
while the complete equation is
\begin{eqnarray}
  2QK\delta n = -ih^2\int\!\dd L\, \rh(L)\rh(Q-L)\; \delta [ &&\!\!
  (1+n(\ell_0)) (1+n(q_0-\ell_0)) n(q_0) \nn&& -
  n(\ell_0)n(q_0-\ell_0) (1+ n(q_0))].
  \label{Boltzmannregime}
\end{eqnarray}
This is, indeed, a linearized Boltzmann equation.

This expression is actually zero, since $Q^2=m^2,\, L^2=m^2$ and
$(Q-L)^2=m^2$ cannot be true at the same time. In our model,
therefore, there is no one-loop correction to the Boltzmann equation.
To get a non-zero right-hand side, we can consider a model with different
masses where $m_1>m_2+m_3$ is true, or we go to higher loops in the
present model.

\subsection{The normal cut region}
\label{sec:normalcut}

Let us now repeat the same analysis for the cut region, where
$k_0>2q_0$ and the first and the third differential equations of
\eqref{sumanddiff} are relevant. First, we take the equation for
$g_{ra}$
\begin{equation}
  \left(Q^2 + \frac{K^2}4-m^2\right) 2g_{ra} = \pi_{ar}\left(
  G^-_{ra}+G^+_{ra}\right) + \left(\Pi^+_{ar}+\Pi^-_{ar}\right)
  g_{ra}.
\end{equation}
In this region the following approximation is valid
\begin{equation}
  G^\pm_{ra}= \frac1{Q^2+K^2/4-m^2 \pm QK \pm ik_0\ep} =
  \frac{\pm 1}{QK+ik_0\ep} +\c O(Q^2+\frac{K^2}4-m^2),
\end{equation}
and so the sum $G^-_{ra}+G^+_{ra}\approx0$. The equation is
approximately homogeneous in $g_{ra}$; we can therefore assume, as
before, that $g_{ra}=g_{ar}=0$.

Then the equation for $g_{rr}$ reads
\begin{equation}
  \left(Q^2 + \frac{K^2}4-m^2\right) 2g_{rr} =\pi_{aa}
  \left(G^-_{ar}+G^+_{ra}\right) +\pi_{ar}G^-_{rr} +\pi_{ra}G^+_{rr}
  + \left(\Pi^+_{ar}+\Pi^-_{ra}\right) g_{rr}.
\end{equation}
We now approximate $G_{rr}$ for small $Q^2+K^2/4-m^2$ and for
$k_0/2>q_0$ as
\begin{eqnarray}
  iG^\pm_{rr}&& =2\pi\sgn(q_0\pm\frac{k_0}2)
  \left(\frac12+n(q_0\pm\frac{k_0}2)\right) \delta(Q^2+\frac{K^2}4-m^2
  \pm QK) \nn && \approx 2\pi\sgn(k_0) \left(\frac12 +
  n(\frac{k_0}2\pm q_0)\right) \delta(QK)
\end{eqnarray}
and, according to the previous approximation:
\begin{equation}
  G^-_{ar}+G^+_{ra}\approx \frac{-1}{QK-ik_0\ep} +\frac{1}{QK+ik_0\ep}
  = -2\pi i\sgn(k_0)\delta(QK).
\end{equation}
So finally all known terms are explicitly proportional to
$\sgn(k_0)\delta(QK)$, and we can have, similarly to the Boltzmann
case, an ansatz that now reads as
\begin{equation}
  ig_{rr}(Q,K) = 2\pi\sgn(k_0) \delta(QK) f(Q,K).
  \label{cutansatz}
\end{equation}
This agrees with the solution of the difference of the SD equations,
if we have considered it as a constraint; but now it is just an ansatz.

In the small $\k$ region $\delta(QK) \approx 1/|k_0| \delta(q_0)$,
we can therefore substitute $q_0\approx 0$. Then the only non-zero
momentum components are $k_0$ and $\q$, and we collect them in a
single four-vector $S=(k_0/2,\q)$. Using spatial rotation symmetry,
$\Pi_{ra}(Q-K/2) = \Pi_{ar}(S)$, and we can write
\begin{equation}
  (S^2-m^2) 2f = \pi_{aa} +(\pi_{ar}+\pi_{ra})
  \left(\frac12+n(s_0)\right) + 2\Pi_{ar}(S) f.
  \label{cutregime}
\end{equation}
This is a linear but non-local wave equation for $f$. On the
right-hand side we have to use the fact that $Q^2+K^2/4-m^2$ is small,
i.e.  $S^2\approx m^2$.

In our concrete model, at one-loop order
$\pi_{aa}(k_0/2,\q)=\Im\Pi_{ar}(S)=0$ at $S^2=m^2$. The second
equality was proved before. In order to prove the first one, we notice
that, according to the ansatz:
\begin{equation}
  \pi_{aa}(Q,K) = -2ih^2\int\!\dd L\, \rh(Q-L) \left(\frac12 +
  n(q_0-\ell_0)\right) \sgn(k_0) 2\pi\delta(LK) f(L,K).
\end{equation}
When $\k=q_0=0$ then $\delta(2LK)\sim\delta(\ell_0)$, and thus
$\rh(Q-L)$ would force $Q-L$ on the mass shell with $q_0-\ell_0=0$;
which is impossible. Thus, at the one-loop order, what remains in
\eqref{cutregime} (writing the $\c P$ dependence explicitly) is
\begin{equation}
  (S^2-m^2) f = \Pi^0_{ar}(S) f -\left(\frac12 + n(s_0)\right)
  \frac{h^2}{s_0} \int\!\frac{d^3\ell}{(2\pi)^3}\, 
  \frac1{(\s-\ell)^2+m^2} f(\ell,s_0) + \left(\frac12 +
  n(s_0)\right)\c P(2s_0).
\label{cuteq}
\end{equation}

We see that the treatment of the two regimes (the Boltzmann and the
normal cut regions) is quite symmetric; the only difference comes from
whether we consider the sum or the difference of the SD equations as
the relevant dynamical differential equation. In the IR sensitive
regions we have for the solution an ansatz that satisfies the other
equation as a constraint.

\subsection{Relaxation time approximation}

Relaxation time approximation for Boltzmann equations is well-known.
It means that we approximate the right-hand side of
\eqref{Boltzmannregime} as $-2iq_0\Gamma_B\delta n$. If this
approximation is valid, then $\delta n(t) \sim \exp(-\Gamma_B t)$. In
our illustrative model, as we have seen, $\Gamma_B=\c O(h^4)$, since
it is zero at the one-loop level.

This approximation scheme can be extended also to the normal cut
regime.  We assume that the operators acting on the right-hand side of
the relevant eq.~\eqref{cuteq} can be represented as a diagonal linear
operator. After rearrangement on the left-hand side, modified
parameters (damping rate \emph{and} mass) will appear. Then
\eqref{cuteq} reads as
\begin{equation}
  \left[\left(\frac{k_0+i\Gamma_c}2\right)^2 -\q^2 -m_{eff}^2\right] f =
  \left(\frac12 + n(\frac{k_0}2)\right)\c P(K).
\end{equation}
At one-loop order the imaginary part $\Gamma_c=0$ , just like in the
Boltzmann case; but, in general, it is non-zero at higher loops. Its
solution is almost the same as the one-loop level result, just the
finite imaginary part requires some care. The ansatz \eqref{cutansatz}
yields
\begin{equation}
  ig_{rr}(Q,K) = \frac{2\pi}{k_0}\delta(q_0) \left(\frac12 +
  n(\frac{k_0}2)\right) \frac1{(k_0+i\Gamma_c)^2/4 -\q^2  -m_{eff}^2}
  \c P(k_0).
\end{equation}
When compared with the definition $ig_{rr}(Q,K)=R(Q,K)\c P(K)/2$, we
find for $\int\dd Q\,iR$, after evaluating the $q_0$ integral,
\begin{equation}
  \int\dd Q\, iR(Q,K)= \frac1{\pi^2 k_0} \left(\frac12 +
  n(\frac{k_0}2)\right) \int\limits_0^\infty\!dq\,
  \frac{q^2}{(k_0+i\Gamma_c)^2/4-m_{eff}^2 -\q^2 }.
\end{equation}
After performing the integration and dropping the infinite piece (mass
renormalization), we finally arrive at
\begin{equation}
  \int\dd Q\, i R(Q,K) = \frac i{4\pi k_0} \left(1+ 2n(\frac{k_0}2)
  \right) \sqrt{\frac{(k_0+i\Gamma_c)^2}4 - m_{eff}^2}.
\end{equation}
The discontinuity, calculated as twice the imaginary part, reads
\begin{eqnarray}
  \int\dd Q\, \mathop{\mathrm{Disc}}_{k_0} iR(Q,K) && = \frac1{4\pi
  k_0} \left(1 + 2n(\frac{k_0}2) \right)
  \,2\Re\sqrt{\frac{(k_0+i\Gamma_c)^2}4 - m_{eff}^2} \nn&& = 
  \frac1{4\pi} \left(1 + 2n(\frac{k_0}2) \right)\; \frac2{k_0}\,
  F\left(\frac{k_0^2}4 - M^2, \frac{k_0\Gamma_c}2 \right)
  \label{result}
\end{eqnarray}
where $M^2 = m_{eff}^2+\Gamma_c^2$ and
\begin{equation}
  F(x,y) = \frac1{\sqrt{2}} \left(x + \sqrt{x^2+y^2}\right)^{1/2}.
  \label{F}
\end{equation}
Let us first analyse some properties of this result
\begin{itemize}
\item If $\Gamma_c=0$ then, because of $F(x,y=0) = \Theta(x)\sqrt{x}$,
  we get back the normal cut part of \eqref{appDiscR} with the
  substitution $m\to M$.
\item Using perturbation theory without resummation, we see IR
  divergences (coming from higher loops) if $m=0$. This is because
  $M^2\sim h^2$, thus we can compute power series in $M^2$, which
  yields
  \begin{equation}
    \frac{2F}{k_0}\biggr|_{pert} = \frac{2}{k_0}
    \sqrt{\frac{k_0^2}4-M^2} = 1 - \frac{2M^2}{k_0^2} + \c O(h^4).
  \end{equation}
\item For $M^2=0$
  \begin{equation}
    \frac{2F}{k_0}\biggr|_{M^2=0} = \frac1{\sqrt{2}}\left( 1 + \sqrt{
    1+ \frac{4\Gamma_c^2}{k_0^2}} \right)^{1/2}
    \stackrel{k_0\to 0}{\longrightarrow} \sqrt{\frac{\Gamma_c}{k_0}}
  \end{equation}
  yields a divergent (although integrable) behaviour.
\item At $k_0=0, M\neq 0$ we get
  \begin{equation}
    \frac{2F}{k_0}\biggr|_{k_0\to 0} = \frac{\Gamma_c}{4M} \sgn(k_0).
  \end{equation}
  This is a finite (and non-analytic) result.
\item While the free result exhibited a non-analytic square-root
  behaviour near the threshold, higher-loop corrections smoothen out
  this non-analyticity. A non-analytic behaviour implies power-law
  real time dependence \cite{oneloopcalc,background-time}, which now
  acquires an exponential damping multiplicative correction, just like
  the solution of the Boltzmann equation. On the other hand, at
  $k_0=0$, there appears a new non-analytic contribution.
\end{itemize}

\begin{figure}[htbp]
  \begin{center}
    \includegraphics[height=8cm,angle=270]{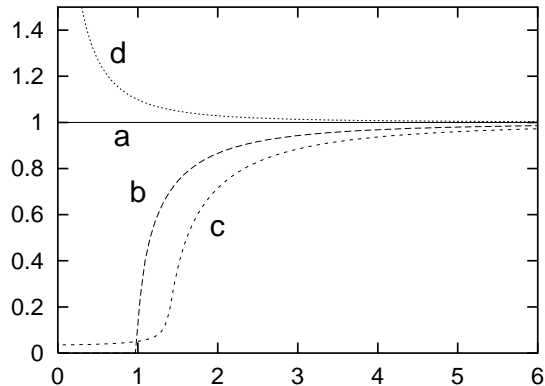}
    \caption{The function $F/k_0$ (cf. \eqref{F} for different
      parameters: \emph{a:} $M=\Gamma_c=0$, \emph{b:} $M=1,\,
      \Gamma_c=0$, \emph{c:} $M=1.4,\,\Gamma_c=0.1$, \emph{d:} $M=0,\,
      \Gamma_c=0.1$.}
    \label{fig:gorbek}
  \end{center}
\end{figure}

In order to see these effects we plotted the function $2F/k_0$ in
Fig.~\ref{fig:gorbek}.
Curve \emph{a} shows the $M=\Gamma_c=0$ result; \emph{b} is the free
result with $m=1$; \emph{c} shows how it changes at higher loops: it
acquires a finite shift in the mass as well as a damping that smears
out the threshold non-analyticity (in the plot we have used $M=1.4,\,
\Gamma_c=0.1$). Finally curve \emph{d} shows the $M=0,\, \Gamma_c\neq
0$ case.

\section{Comments on the fermionic case}
\label{sec:fermcomments}

\begin{figure}[htbp]
  \begin{center}
    \includegraphics[height=8cm,angle=270]{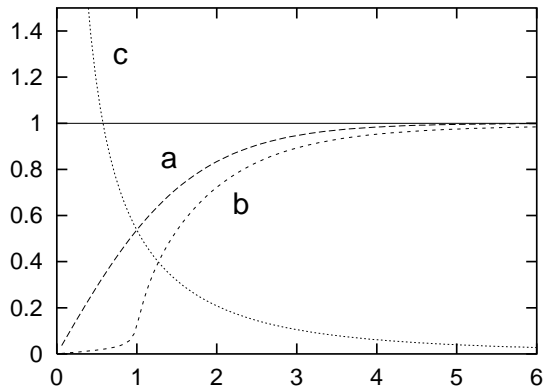}
    \caption{Expected result of
      $\exvs{[j_\mu,j^\mu]}(k_0)/k_0^2$. Curve \emph{a} is the one
      loop result, curve \emph{b} is the result after resummation.
      Curve \emph{c} indicates the size of the 2-loop corrections.}
    \label{fig:ferm}
  \end{center}
\end{figure}

Let us make some comments on the fermionic case. In our simple model,
in relaxation time approximation, the only change in the final result,
compared to the free case, was a modification just in the threshold
function $\Theta(x)\sqrt{x}\to F(x,y)$, cf. \eqref{result}. This seems
to be the consequence of the relaxation time approximation, not
depending on the specific model, so we may assume that this feature
generalizes to other theories, too.  In case of dilepton production we
have a fermionic system with approximately zero masses. The rate of
the dilepton production is proportional to $\exvs{[j_\mu,j^\mu]}(K)$
\cite{dilepton}; when computed perturbatively, this is very similar to
our observable $\exvs{[\ph^2,\ph^2]}(K) = \int\dd Q\,
\mathop{\mathrm{Disc}}_{k_0} iR(Q,K)$. At $\k=0$ the production rate
is $\exvs{[j_\mu,j^\mu]}(k_0)/k_0^2 \sim \tanh(\beta k_0/4)$; if we
assume the same behaviour for the higher loop pinch singularities in
the fermionic theory as they appear in our model, we obtain changes in
the production rate qualitatively similar to Fig. \ref{fig:ferm}.
Here curve \emph{a} shows the free result, curve \emph{b} shows the
expected result of the resummation, applying the present analysis.
Curve \emph{c} indicates the size of the 2-loop perturbative
corrections. This figure suggests that, although the two-loop
correction is divergent in the $k_0\to 0$ limit, the resummed result
can be convergent, or even zero. With the present approximation the
discontinuity goes linearly to zero for vanishing momentum, and its
slope is $\sim \Gamma_c/(4TM)$. Our approximate result is quite
reminiscent of the numerical (MEM method) findings of Karsch et {\it
  al.}  \cite{Karschetal}.

\section{Conclusion}
\label{sec:conclusion}

Let us summarize the main results of the paper:
\begin{itemize}
\item \emph{Refinement of the conventional treatment of Kadanoff--Baym
    equations.} Starting from the pair of non-equilibrium
  Schwinger--Dyson equations, applying Wigner transformation and
  taking their sum and difference, we arrive at the KB equations. The
  conventional interpretation is the following: for mild variations of
  the average coordinates, one of the equations is understood as a
  conventional first-order partial differential equation: the quantum
  Boltzmann equation. In the other equation one neglects the
  dependence on the average coordinates, as a consequence of which it
  becomes a constraint.
  
  We propose a somewhat different approach. Based on the thorough
  analysis of the one-loop contribution, and resumming the IR (pinch)
  singularities arising from higher loops, we argue that \emph{both}
  equations have to be interpreted as dynamical differential
  equations, each one beeing relevant in complementary kinematical
  regimes.  The complete solution is the sum of the two equations.
  Concentrating on the pinch-singular contributions, we suggested for
  both solutions an ansatz that satisfies the constraint equation of
  the conventional interpretation. In this way one of the equations is
  still a (quantum) Boltzmann equation; the other one, however,
  represents a new contribution, reproducing in a resummed form the
  particle production cut.
\item \emph{Features of the new contribution.} The new contribution is
  responsible at tree level for the appearance of the normal cut; it
  is IR (pinch) singular at higher loops. We have constructed the
  relevant equation near the IR-singular regime (see
  \eqref{cutregime}).  It is a wave equation for the quantity that is
  the analogue of the particle number variation in the Boltzmann case.
\item \emph{Solution in the relaxation time approximation.} We can try
  to diagonalize the linear differential equations and keep the
  eigenmode with the smallest damping rate: this is the relaxation
  time approximation in effect. In the Boltzmann regime this procedure
  changes the infinitesimal $i\ep$ term of the Landau prescription to
  the finite imaginary part $i\Gamma_B$, associated with the presence
  of an effective term in the equation of motion, which is linear in
  the time derivative. In the normal cut regime there are two
  contributions: a real one, which is in fact a mass shift, and an
  imaginary one, which is analogous to the damping rate in the
  Boltzmann case. Computing the normal cut contribution with this
  approximation we can observe two effects: the shift of the threshold
  location and the smoothing of the threshold non-analyticity. If
  treated in finite orders of perturbation theory, both effects may
  lead to divergences.
\item \emph{Prospects for generalization.} The present computation was
  performed for the case of the cubic scalar theory, and direct
  computations in other more relevant models are required before one
  can have quantitative predictions: this is a project for the future.
  Still, some findings of the present calculation seem to be robust
  enough to conjecture similar features to be valid in other cases as
  well. The ``dual'' interpretation of the KB equations, the wave
  equation form of the equation describing the cut contributions, the
  imaginary \emph{and} real parts in the relaxation time approximation
  probably generalize to the fermionic case as well. Then we
  conjecture for the quantity $\exvs{[j_\mu,j^\mu]}(k_0)/k_0^2$,
  relevant to the determination of the dilepton production rate, the
  behaviour indicated on Fig.~\ref{fig:ferm}. While two-loop order
  perturbation theory gives a divergent contribution (curve \emph{c})
  at threshold, the resummed result shows a much milder behaviour
  (curve \emph{b}). This qualitative prediction displays remarkable
  similarity to recent numerical data \cite{Karschetal}.
\end{itemize}

\section*{Acknowledgements}

The author would like to thank to D. Litim, A. Patk{\'o}s and P.
Petreczky for discussions and proposals. This work was partially
supported by the Hungarian Science Fund (OTKA).

\appendix

\section{Appendix: The formalism}
\label{sec:formalism}

In the Keldysh formalism for a complex scalar field\footnote{For
  fermionic fields the formalism is analogous, except for some
  appropriately placed minus signs, see \cite{LandsvanWeert}.} we
define the propagators
\begin{eqnarray}
  & iG_{11}(t,t') = \Theta(t-t') \exv{\Phi(t)\Phi^\dagger(t')} +
  \Theta(t'-t) \exv{\Phi^\dagger(t')\Phi(t)}, \quad  
  & iG_{12}(t,t') = \exv{\Phi^\dagger(t')\Phi(t)},\nn
  & iG_{22}(t,t') = \Theta(t'-t) \exv{\Phi(t)\Phi^\dagger(t')} +
  \Theta(t-t') \exv{\Phi^\dagger(t')\Phi(t)}, \quad 
  & iG_{21}(t,t') = \exv{\Phi(t)\Phi^\dagger(t')},
\end{eqnarray}
where the expectation value is defined in general as
\begin{equation}
  \exvs{\hat A} = \Tr \hat\rho \hat A,
  \label{exvdef}
\end{equation}
where $\hat \rho$ is the density matrix. Formally we introduce two
types of fields ($\Phi_1$ and $\Phi_2$), which are representations of
the original field on the different branches of the closed time
contour. Their expectation values are defined as
$\exv{\Phi_i\Phi_j}=iG_{ij}$.

These propagators are not independent, there is a relation between
them
\begin{equation}
  G_{11}+G_{22} = G_{12} + G_{21}.
  \label{Grelation}
\end{equation}
In fact, there are only two independent expectation values, e.g.
$G_{12}$ and $G_{21}$, since $G_{11}$ and $G_{22}$ are combinations of
these two and the theta function. In equilibrium even these two are
related through the KMS condition \cite{LandsvanWeert}.

While these propagators determine the free theory, in case of
interactions we have to modify the interaction Lagrangian as (cf.
\cite{LandsvanWeert})
\begin{equation}
  \c L_I(\Phi) \longrightarrow \c L_I(\Phi_1)-\c L_I(\Phi_2).
\end{equation}

In the R/A formalism \cite{RAformalism} we introduce new fields
\begin{equation}
  \Phi_r = \frac{\Phi_1 + \Phi_2}2,\qquad \Phi_a = \Phi_1-\Phi_2.
\end{equation}
Their time-ordered products (denoting $\Phi_i(x)$ by $\Phi_i$ and
$\Phi^\dagger_i(y)$ by $\Phi'_i$) are expressible through the
two-point functions as follows:
\begin{eqnarray}
  && G_{rr} = \frac1{4i} \exv{\Phi_1\Phi'_1+ \Phi_1\Phi'_2 +
  \Phi_2\Phi'_1 + \Phi_2\Phi'_2} =  \frac14 (G_{11} + G_{12} + G_{21}
  + G_{22}) = \frac{G_{12} + G_{21}}2,\nn
  && G_{ra} = \frac1{2i} \exv{\Phi_1\Phi'_1 - \Phi_1\Phi'_2 +
  \Phi_2\Phi'_1 - \Phi_2\Phi'_2} = \frac12 (G_{11} - G_{12} + G_{21}
  - G_{22}) = G_{11} - G_{12},\nn 
  && G_{ar} = \frac1{2i} \exv{\Phi_1\Phi'_1 + \Phi_1\Phi'_2 -
  \Phi_2\Phi'_1 - \Phi_2\Phi'_2} =  \frac12 (G_{11} + G_{12} - G_{21}
  - G_{22}) = G_{11} - G_{21},\nn 
  && G_{aa} = -i\exv{\Phi_1\Phi'_1 - \Phi_1\Phi'_2 -
  \Phi_2\Phi'_1 + \Phi_2\Phi'_2} =  G_{11} - G_{12} - G_{21} + G_{22}
  =0.
\end{eqnarray}

In a non-equilibrium situation the propagators depend on the two
space-time coordinates separately, and one can define several integral
representations for the propagators. The double Fourier transform of a
function of two arguments is defined as
\begin{equation}
   f(p,q) = \int\!dxdy\, e^{ipx-iqy} f(x,y),\qquad
  f(x,y) = \int\!\dd p\dd q\, e^{-ipx+iqy}  f(x,y),
\end{equation}
where $\dd{p} = d^D p/(2\pi)^D$ in $D$ dimensions. Note that we define
the momentum of the $y$ variable with a relative negative sign!  We
use the same symbols for the function and its Fourier transform, the
type of the argument defines which function we are dealing with. The
Wigner transform $\bar f$ of a function $f$ with two arguments is
defined as
\begin{equation}
  \bar f(u,X) = f(X+\frac u2, X-\frac u2),\qquad 
  f(x,y) = \bar f(x-y, \frac{x+y}2),
\end{equation}
i.e. it understands the function to depend on the relative and average
coordinates. Fourier transformation with respect to these coordinates
yields
\begin{eqnarray}
  && \bar f(Q,K) = \int\!dudX\,e^{iQu+iKX} \bar f(u,X) =
  \int\!dxdy\,e^{iQ(x-y)+iK(x+y)/2} f(x,y) =  f(Q + \frac
  K2, Q-\frac K2),\nn && f(p,q) = \bar f(\frac{p+q}2, p-q).
\end{eqnarray}
The derivatives with respect to $x$ and $y$ transform as
\begin{eqnarray}
  &&i\d_x f(x,y) \stackrel{Fourier}{\longrightarrow} p f(p,q)
  \stackrel{Wigner}{\longrightarrow} (Q+\frac K2) \bar f(Q,K),\nn
  &&i\d_y f(x,y)\stackrel{Fourier}{\longrightarrow} -q f(p,q)
  \stackrel{Wigner}{\longrightarrow} (-Q+\frac K2) \bar f(Q,K).
\end{eqnarray}

Later on we will use the Schwinger-Dyson equations for propagators. It
is used to resum bubble series
\begin{equation}
  G_{ab} = G^{0}_{ab} + G^{0}_{ac} \Pi_{cd} G^{0}_{db} + \dots =
  G^{0}_{ab} + G^{0}_{ac} \Pi_{cd} G_{db} = G^{0}_{ab} + G_{ac}
  \Pi_{cd} G^{0}_{db}.
  \label{recursiveSD}
\end{equation}
The indices refer to space-time coordinates as well as to internal
indices (e.g. Keldysh indices). To have a differential form we apply
the free time evolution kernel on both sides. In the R/A formalism
these kernels satisfy ``twin'' equations, which differ only in the
variable to which the derivatives are referring to:
\begin{equation}
  (\d_{x,y}^2 + m^2) G_{rr}^{0}(x,y) = 0\quad \mathrm{and}\quad
  (\d_{x,y}^2 + m^2) G_{ra}^{0}(x,y) = (\d_{x,y}^2 + m^2)
  G_{ar}^{0}(x,y) = -\delta(x-y).
\end{equation}
Using these, we find the SD equations:
\begin{equation}
  \begin{array}[b]{ll}
    \displaystyle -(\d_x^2 + m^2) G_{rr} = \int\!dz\, (\Pi_{aa}
    G_{ar} + \Pi_{ar} G_{rr}),\quad & \displaystyle -(\d_y^2 + m^2)
    G_{rr} = \int\!dz\, (G_{ra}\Pi_{aa}  +  G_{rr} \Pi_{ra})\cr
    \displaystyle -(\d_x^2 + m^2) G_{ra} = \delta(x-y) + \int\!dz\,
    \Pi_{ar} G_{ra},\quad & \displaystyle -(\d_y^2 + m^2) G_{ra} =
    \delta(x-y) + \int\!dz\,G_{ra}\Pi_{ar},\cr
    \displaystyle -(\d_x^2 + m^2) G_{ar} = \delta(x-y) + \int\!dz\,
    \Pi_{ra} G_{ar},\quad & \displaystyle -(\d_y^2 + m^2) G_{ar} =
    \delta(x-y) + \int\!dz\, G_{ar} \Pi_{ra}.\cr
  \end{array}
\label{SD-eq}
\end{equation}
Here we have used the fact that $G_{aa}=0$ is consistent with SD
equations only if $\Pi_{rr}=0$. After the Wigner--Fourier
transformations the equations in (\ref{SD-eq}) look a little more
complicated:
\begin{eqnarray}
  &\left((Q+\frac K2)^2-m^2\right) \bar G_{rr}(Q,K) = &\textstyle
  \int\!\dd k\, \biggl[ \bar \Pi_{aa}(Q+\frac k2,K-k) \bar
  G_{ar}(Q-\frac{K-k}2, k) + \nn&&\textstyle\qquad\qquad+\;
  \bar\Pi_{ar}(Q+\frac k2,K-k) \bar G_{rr}(Q-\frac{K-k}2, k)\biggr]\nn
  &\left((Q-\frac K2)^2-m^2\right) \bar G_{rr}(Q,K) = &\textstyle
  \int\!\dd k\, \biggl[ \bar \Pi_{aa}(Q-\frac k2,K-k) \bar
  G_{ra}(Q+\frac{K-k}2, k) + \nn&&\textstyle\qquad\qquad+\;
  \bar\Pi_{ra}(Q-\frac k2,K-k) \bar G_{rr}(Q+\frac{K-k}2, k)\biggr]\nn
  &\left((Q+\frac K2)^2-m^2\right) \bar G_{ra}(Q,K) = &\textstyle
  \horsl{\delta}(K) + \int\!\dd k\, \bar \Pi_{ar}(Q+\frac k2,K-k) \bar
  G_{ra}(Q-\frac{K-k}2, k)\nn
  &\left((Q-\frac K2)^2-m^2\right) \bar G_{ra}(Q,K) = &\textstyle
  \horsl{\delta}(K) + \int\!\dd k\, \bar \Pi_{ar}(Q-\frac k2,K-k) \bar
  G_{ra}(Q+\frac{K-k}2, k)\nn
  &\left((Q+\frac K2)^2-m^2\right) \bar G_{ar}(Q,K) = &\textstyle
  \horsl{\delta}(K) + \int\!\dd k\, \bar \Pi_{ra}(Q+\frac k2,K-k) \bar
  G_{ar}(Q-\frac{K-k}2, k)\nn
  &\left((Q-\frac K2)^2-m^2\right) \bar G_{ar}(Q,K) = &\textstyle
  \horsl{\delta}(K) + \int\!\dd k\, \bar \Pi_{ra}(Q-\frac k2,K-k) \bar
  G_{ar}(Q+\frac{K-k}2, k),
  \label{WigSD}
\end{eqnarray}
where $\horsl{\delta}(p)=(2\pi)^D\delta(p)$ in $D$ dimensions. The
$\pm K$ equations are related to each other by complex conjugation and
$K\to-K$ substitution.

\section{Appendix: Detailed computation of Section~\ref{sec:oneloop}}
\label{sec:compdet}

In the product of delta functions \eqref{proddelta} we observe that
the first delta function $\delta(Q^2+QK+K^2/4 -m^2)$ implies $QK=
m^2-Q^2-K^2/4$, which yields $\delta(Q^2-QK+K^2/4 -m^2)
\to\delta(2QK)$. This, in turn, means that one can set $QK=0$ in its
coefficient, and therefore we can write $\delta(Q^2+QK+K^2/4 -m^2)\to
\delta(Q^2+K^2/4 -m^2)$. We obtain in this way the second form of
\eqref{proddelta}. In order to get the third form the following steps
are done:
\begin{eqnarray}
  D= && \delta(2q_0k_0-2qkx) \delta(q_0^2-q^2+\frac{K^2}4 - m^2) =
  \frac1{2|k_0|} \delta(q_0-\frac{qkx}{k_0}) \delta(q^2
  (\frac{k^2x^2}{k_0^2} -1)+\frac{K^2}4 - m^2) =\nn&&= \frac{|k_0|}2
  \frac 1{|k_0^2-k^2x^2|} \delta(q_0-\frac{qkx}{k_0})\delta(q^2 -
  \frac{k_0^2}4\, \frac {K^2-4m^2}{k_0^2-k^2x^2}) \nn&& =
  \frac{|k_0|}2\frac 1{|k_0^2-k^2x^2|} \delta(q_0-\frac{qkx}{k_0})
  \frac1{2q} \left[ \delta(q-\frac{k_0}2{\c Z})
    +\delta(q+\frac{k_0}2{\c Z}) \right] \nn&& =
  \frac1{2\,{\c Z}\,|k_0^2-k^2x^2|} \delta(q_0-\frac{kx}2
  \sgn(k_0){\c Z}) \delta(q-\frac{|k_0|}2{\c Z}).
\end{eqnarray}
In the third line we have introduced the quantity ${\c
  Z}=\sqrt{(K^2-4m^2)/({k_0^2-k^2x^2})}$. In order to get the fourth
line we used the fact that, since $q$ is positive, only one of the two
delta functions appearing on the third line can be satisfied,
therefore $q=|k_0|{\c Z}/2$. The expression is non-zero if $\c Z$ is
real, i.e.  when $(K^2-4m^2)/({k_0^2-k^2x^2})>0$.  To examine this
condition we split the $K^2$ range into two regions:
\begin{itemize}
\item $K^2>4m^2$; then $k_0^2>k^2>k^2x^2$ (since $|x|<1$), so the
  numerator and the denominator are positive;
\item $K^2<4m^2$; then the numerator is negative, the denominator thus
  has to be negative as well. This yields $k_0^2<k^2x^2$, which
  implies $1>|x|>|k_0|/k$ and $k^2>k_0^2$, i.e. $K^2<0$.
\end{itemize}
Therefore $\c Z$ is real if $K^2>4m^2$ or $K^2<0$.

We can examine the relative size of $|q_0|$ and $|k_0|$ from the last
form of \eqref{proddelta}. To find it we rewrite $\c Z$ as
\begin{equation}
  {\c Z}^2 = \frac{K^2-4m^2}{k_0^2-k^2x^2} = 1 -
  \frac{k^2(1-x^2)+4m^2}{k_0^2-k^2x^2}.
\end{equation}
The numerator is always positive, thus
\begin{itemize}
\item if $K^2>4m^2$, then $|x|\le1,\;\c Z\le1$, and one has $|q_0|
  =\frac{|kx|}2{\c Z} \le \frac k2 \le \frac{|k_0|}2$;
\item if $K^2<0$ then $1\ge|x|\ge|k_0|/k,\; \c Z\ge 1$, giving $|q_0|
  =\frac{|kx|}2{\c Z} \ge \frac{|k_0|}2$
\end{itemize}

For the $Q$-integration appearing in \eqref{DiscR:qkform}
\begin{eqnarray}
   && \mathop{\mathrm{Disc}}_{k_0} iR(Q,K) =\frac{4 \pi^2}
  {\sqrt{(k_0^2-k^2x^2) (K^2-4m^2)}}\delta(q_0-\frac{kx}2
  \sgn(k_0){\c Z}) \delta(q-\frac{|k_0|}2{\c Z}) \times\nn&&
  \left[\Theta(K^2-4m^2) \left(1+ n(\frac{k_0}2-q_0)+
  n(\frac{k_0}2+q_0)\right) + \Theta(-K^2) \left(n(q_0-\frac{k_0}2) -
  n(q_0+\frac{k_0}2)\right) \right]
\end{eqnarray}
we write the integration measure as
\begin{equation}
  \frac1{8\pi^3} \int \!dq_0\,dq\,dx\,q^2.
\end{equation}
The following change of variable is made:
\begin{equation}
  x\to y = \frac{kx}2{\c Z},\quad \frac{dy}{dx} = \frac{k_0^2 k}2
  \sqrt{\frac{K^2-4^2} {(k_0^2-k^2x^2)^3}},\quad 
  y\in\frac12 \left\{ \begin{array}[c]{ll}
      [-k{\c Z}_1, \,k{\c Z}_1],\quad&\mathrm{if}\; K^2>4m^2\cr
      [-\infty,-k{\c Z}_1] \cup [k{\c Z}_1,\infty],\;&\mathrm{if}\;
      K^2<0\cr 
    \end{array}\right.
\end{equation}
where ${\c Z}_1 = {\c Z}(x=1) =\sqrt{1- 4m^2/K^2}$. Then we have
\begin{equation}
  \frac1{8\pi^3} \int\limits_{-\infty}^\infty\!dq_0
  \int\limits_0^\infty \!\!\!dq \left[ \Theta(K^2-4^2)\!\!\!\!
  \int\limits_{-k{\c Z}_1/2}^{k{\c Z}_1/2}\!\!\!dy
  +\Theta(-K^2)\!\!\!\! \int\limits_{-\infty}^{-k{\c Z}_1/2}\!\!\!dy +
  \Theta(-K^2)\!\!\!\! \int\limits_{k{\c Z}_1/2}^{\infty}\!\!\!dy
  \right]\;\frac{2q^2}{k_0^2 k} \,\sqrt{
  \frac{(k_0^2-k^2x^2)^3}{K^2-4m^2}}. 
\end{equation}
Putting the ingredients together we find the result quoted in
\eqref{DiscR:Qintform}:
\begin{eqnarray}
  \int\!\dd Q\, \mathop{\mathrm{Disc}}_{k_0} iR(Q,K) =
  \frac{\sgn(k_0)}{8\pi k} \biggl[ && \Theta(K^2-4m^2)
  \int\limits_{-k{\c Z}_1/2}^{k{\c Z}_1/2}\!dy\, \left( 1 +
    n(\frac{|k_0|}2-y) +n(\frac{|k_0|}2+y)\right) +\nn&&  2\Theta(-K^2)
  \int\limits_{k{\c Z}_1/2}^\infty\! dy\, \left( n(y-\frac{|k_0|}2) -
    n(y+\frac{|k_0|}2) \right)\biggr].
\end{eqnarray}


\begin{thebibliography}{99}
  
\bibitem{Karschetal} F. Karsch, E. Laermann, P. Petreczky, S. Stickan
  and I. Wetzorke, {\em Phys. Lett.} {\bf B530}, 147 (2002)
\bibitem{AltherrAurenche} T. Altherr and P. Aurenche, \emph{Z. Phys.}
  \textbf{C45}, 99 (1989)
\bibitem{BraatenPisarski} E. Braaten, R. D.  Pisarski and T. Ch. Yuan,
  \emph{Phys.  Rev. Lett.} \textbf{64}, 2242 (1990)
\bibitem{Aurenche} P. Aurenche, F. Gelis and H. Zaraket, \emph{Phys.
    Rev.} \textbf{D61}, 116001 (2000), P. Aurenche, F.~Gelis and H.
  Zaraket, \emph{Phys. Rev.} \textbf{D62}, 096012 (2000)
\bibitem{AMY} P. Arnold, G. D. Moore and L. G. Yaffe, \emph{JHEP}
  \textbf{11}, 057 (2001); P. Arnold, G. D. Moore and L. G. Yaffe,
  \emph{JHEP} \textbf{12}, 009 (2001); P. Arnold, G. D. Moore and L. G.
  Yaffe, hep-ph/0204343
\bibitem{BVW} D. Boyanovsky, H. J. de Vega and S.-Y. Wang, \emph{Phys.
    Rev.} \textbf{D61}, 065006 (2000).
\bibitem{LandsvanWeert} N. P. Landsmann and Ch. G. van Weert,
  \emph{Phys. Rep.} \textbf{145}, 141 (1987)
\bibitem{pinch-sing} T. Altherr and D. Seibert, \emph{Phys. Lett.}
  \textbf{B333}, 149 (1994)
\bibitem{AJ} A. Jakov\'ac, \emph{Phys. Rev.} \textbf{D65}, 085029
  (2002)
\bibitem{dilepton} J. Kapusta, \emph{Phys. Lett.}  \textbf{B136}, 201
  (1984); L. D. McLerran and T. Toimela, \emph{Phys. Rev.}
  \textbf{D31}, 545 (1985)
\bibitem{oneloopcalc} A. Jakov\'ac, A. Patk{\'o}s, P. Petreczky and
  Zs. Sz{\'e}p, \emph{Phys. Rev.} \textbf{D61}, 025006 (1999)
\bibitem{background-time} D. Boyanovsky, I. D. Lawrie and D. S. Lee
  \emph{Phys. Rev.} \textbf{D54}, 4013 (1996); D. Boyanovsky, M.
  D'Attanasio, H. J. de Vega and R. Holman \emph{Phys.  Rev.}
  \textbf{D54}, 1748 (1996).
\bibitem{RAformalism} E. Wang and U. Heinz, \emph{Phys. Lett.}
  \textbf{B471}, 208 (1999); K.-C.  Chou, Z.-B. Su, B.-L. Lao and L.
  Yu, \emph{Phys. Rep.} \textbf{118}, 1 (1985)
\end{thebibliography}
\end{document}